\documentclass{article}

\usepackage{PRIMEarxiv}

\usepackage[utf8]{inputenc} 
\usepackage[T1]{fontenc}    
\usepackage{hyperref}       
\usepackage{url}            
\usepackage{booktabs}       
\usepackage{amsfonts}       
\usepackage{nicefrac}       
\usepackage{microtype}      
\usepackage{lipsum}
\usepackage{fancyhdr}       
\usepackage{graphicx}       
\graphicspath{{media/}}     
\usepackage{authblk}
\usepackage{tipa}

\title{Xiwu: A Basis Flexible and Learnable LLM for High Energy Physics
}

\author[1]{\textbf{Zhengde~Zhang}}
\author[1]{\textbf{Yiyu~Zhang}}
\author[1]{\textbf{Haodong~Yao}}
\author[2]{\textbf{Jianwen~Luo}}
\author[1]{\textbf{Rui~Zhao}}
\author[1]{\textbf{Bo~Huang}}
\author[3,1]{\textbf{Jiameng~Zhao}}
\author[1]{\textbf{Yipu~Liao}}
\author[1]{\textbf{Ke~Li}}
\author[1]{\textbf{Lina~Zhao}}
\author[1]{\textbf{Jun~Cao}}
\author[1,\thanks{\textit{\underline{Corresponding Author}}: 
    \textbf{qfz@ihep.ac.cn}}]{\textbf{Fazhi~Qi}}
\author[1,\thanks{\textit{\underline{Corresponding Author}}: 
    \textbf{yuancz@ihep.ac.cn}}]{\textbf{Changzheng~Yuan}}

\affil[1]{Institute of High Energy Physics, Chinese Academy of Sciences, Beijing 100049, China}
\affil[2]{School of Physical Sciences, University of Chinese Academy of Sciences, Beijing 100049, China}
\affil[3]{School of Computer and Artificial Intelligence, ZhengZhou University, Henan 450066, China}

\begin{document}
\maketitle

\begin{abstract}
Large Language Models (LLMs) are undergoing a period of rapid updates and changes, with state-of-the-art (SOTA) model frequently being replaced. When applying LLMs to a specific scientific field, it's challenging to acquire unique domain knowledge while keeping the model itself advanced. To address this challenge, a sophisticated large language model system named as Xiwu has been developed, allowing you switch between the most advanced foundation models and quickly teach the model domain knowledge. In this work, we will report on the best practices for applying LLMs in the field of high-energy physics (HEP), including: a seed fission technology is proposed and some data collection and cleaning tools are developed to quickly obtain domain AI-Ready dataset; a just-in-time learning system is implemented based on the vector store technology; an on-the-fly fine-tuning system has been developed to facilitate rapid training under a specified foundation model. \\
The results show that Xiwu can smoothly switch between foundation models such as LLaMA, Vicuna, ChatGLM and Grok-1. The trained Xiwu model is significantly outperformed the benchmark model on the HEP knowledge Q\&A and code generation. 
This strategy significantly enhances the potential for growth of our model's performance, with the hope of surpassing GPT-4 as it evolves with the development of open-source models. 
This work provides a customized LLM for the field of HEP, while also offering references for applying LLM to other fields, the corresponding codes are available on Github \href{https://github.com/zhangzhengde0225/Xiwu}{https://github.comzhang/zhengde0225/Xiwu}. 
\end{abstract}

\begin{center}
    \noindent\includegraphics[width=0.7\linewidth]{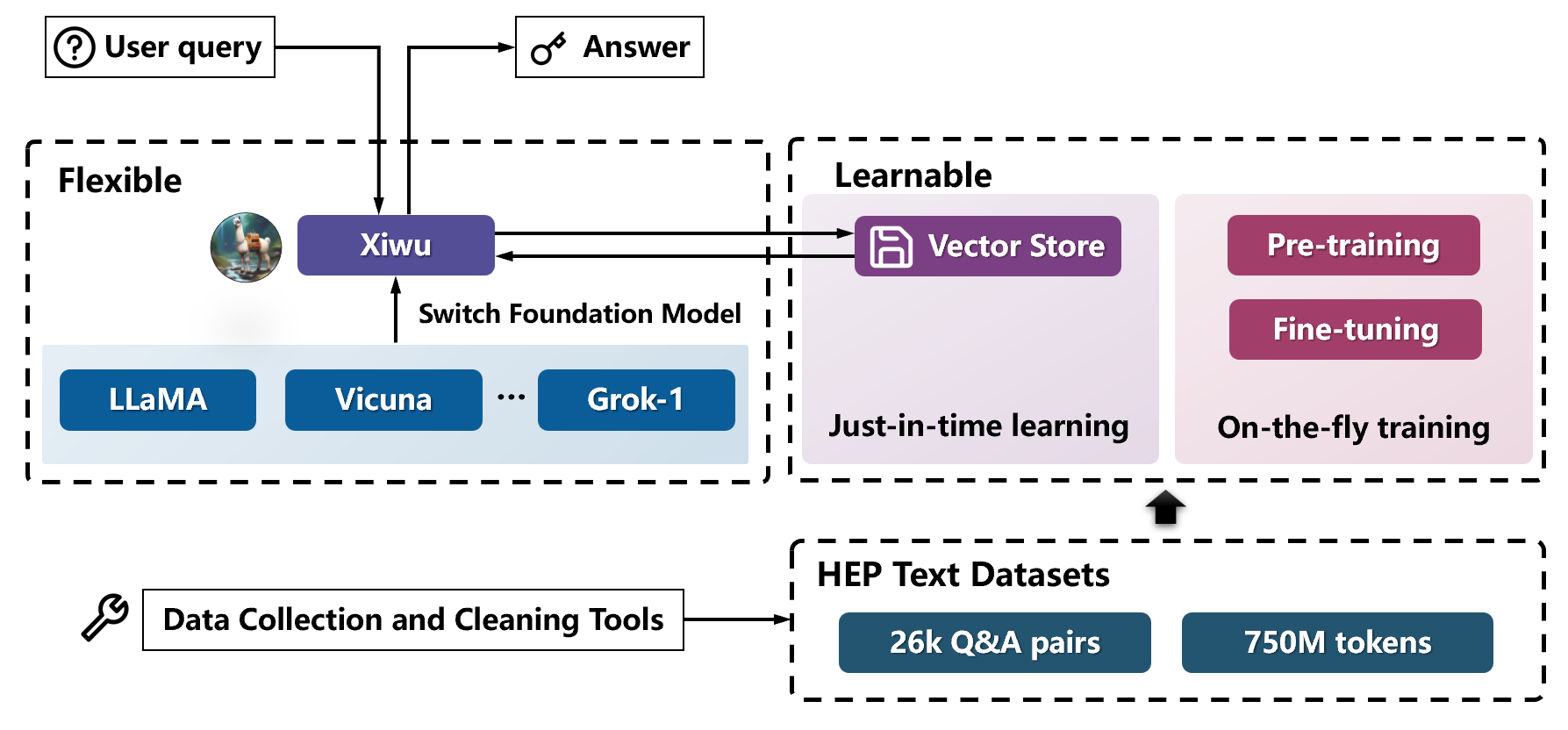}
\end{center}

\keywords{LLM \and Deep Learning \and Artificial Intelligence \and Particle Physics \and HEP}

\section{Introduction}

\begin{figure}
  \centering
  \includegraphics[width=0.95\linewidth]{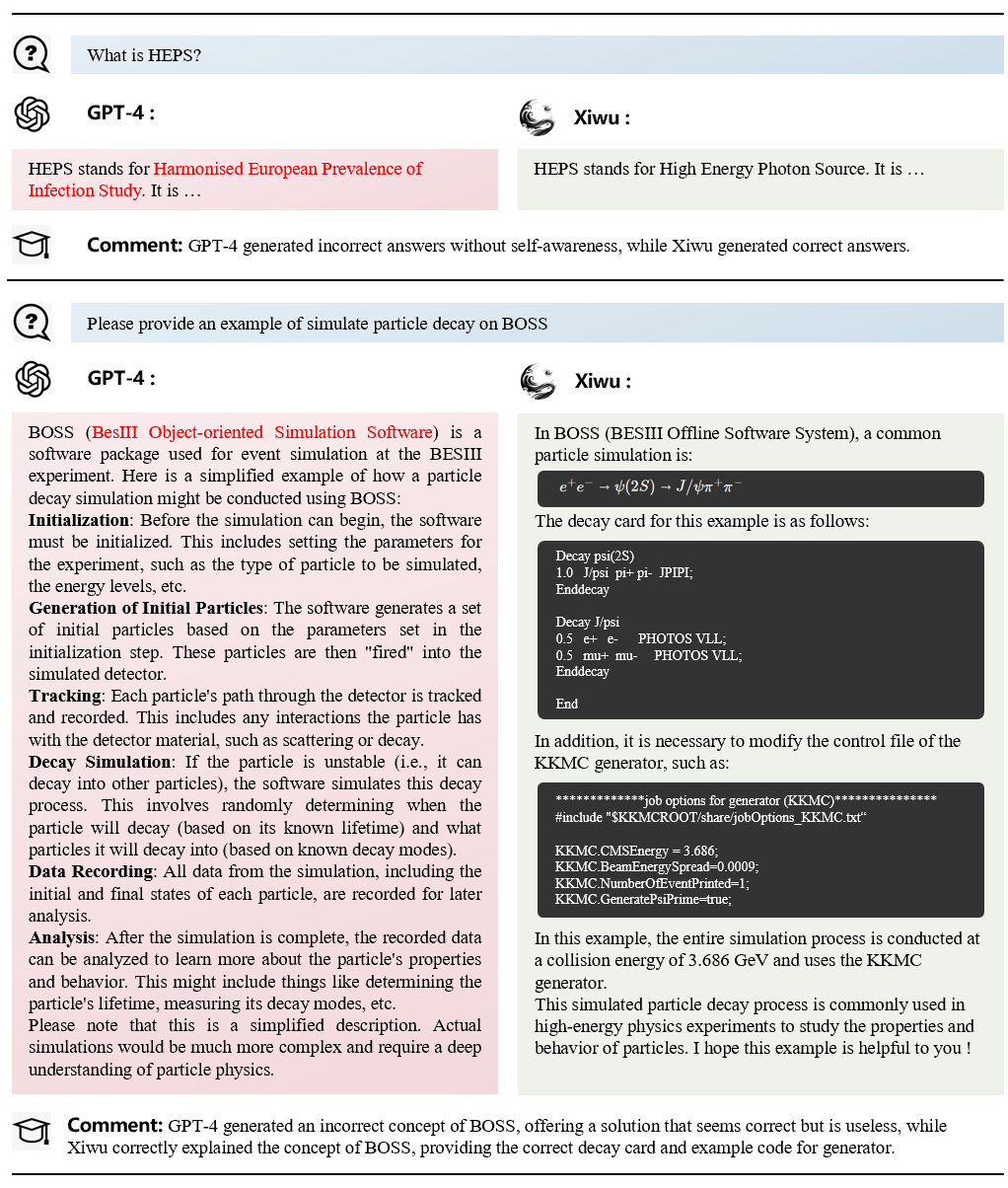}
  \caption{The hallucination of GPT-4 when answering domain questions.}
  \label{fig:hallucination}
\end{figure}

\paragraph{Large Language Models} The Large Language Models (LLMs) such as GPT-4~\cite{achiam2023gpt} and LLaMA\cite{touvron2023llama, touvron2023llama2} have exhibited capabilities beyond expectations in terms of general intent understanding, robust continuous dialogue, intelligent interaction correction, and adequate logical reasoning, actively propelling the shifts towards an intelligent paradigm in scientific research.
\paragraph{High Energy Physics} High Energy Physics (HEP) is a crucial branch of physics that delves into the characteristics and interactions of the most basic elements of matter. It serves as an essential field for unraveling the universe's underlying principles and laws. Creating Large Language Models (LLMs) for High Energy Physics (HEP) can play a pivotal role in streamlining research activities, including literature review, coding, data management, analysis of physics, interpretation of results, and manuscript composition. This advancement can free researchers from mundane tasks that demand less creativity, thereby amplifying the efficiency and productivity of scientific investigations.
\paragraph{Hallucinations} However, the LLMs can suffer from \textbf{hallucinations}, i.e. “produce content that is nonsensical or untruthful in relation to certain sources”~\cite{radford2019language, radford2018improving}, which is particularly harmful in scientific fields. As shown in Figure \ref{fig:hallucination}, even the most powerful LLM (GPT-4) can struggle with domain-specific questions or generating scientific code of HEP. The capability of general LLMs to precisely tackle specialized topics or develop code for focused areas may be constrained.
\paragraph{Way to Reduce Hallucinations} There are two methods can enhance the model's ability to handle domain-specific tasks: one is further training the model, including both secondary pre-training and fine-tuning, and the other is Retrieval-Augmented Generation (RAG). The foundation models are often referred to as L0 (Level 0) models, such as LLaMA and GPT-3. Models fine-tuned for specific tasks based on the foundation model are called L1 models, such as LLaMA-Chat, Vicuna~\cite{zheng2023judging} and GPT-3.5~\cite{ChatGPT2022}, which is further adjusted for conversational tasks. Models that are further fine-tuned with domain-specific knowledge can be termed as L2 models. 
\paragraph{HEP·Xiwu LLM} Here, we introduce the \textbf{Xiwu} (\textipa{/SeI"wu:/}), a L2 large language model that exhibits fewer hallucinations in HEP field. 
The most significant features of this model are its basis flexibility and learnability. Basis flexible means that the foundational model can be changed, becoming increasingly powerful with the upgrade of open-source models. Learnable implies that once the foundational model is updated, it can quickly be taught domain knowledge. Specifically, a just-in-time learning system based on a vector store has been implemented, allowing for instant knowledge updates to the Xiwu at a low cost.
\paragraph{Why Consider Flexible?} Training a foundation LLM is extremely costly. For a specific scientific field, on one hand, training LLM from scratch is unaffordable; on the other hand, even if trained LLM from scratch, it might not keep up with the pace of development of open-source models. "Using the strengths of others to improve oneself," we believe that having a flexible foundational model is the best strategy to maintain the most advanced performance of the model at the lowest cost.
\paragraph{Contributions} The contributions are as follows:
\begin{enumerate}
    \item The seed fission technology is proposed and realized, which has proven its effectiveness in acquiring training Q\&A data related to the field with only one seed. It can be widely applied in various fields.
    \item A basis flexible LLM system has been strategized and implemented, evolving from the initial LLaMA to now support upgrades to LLaMA2 and Vicuna, and beyond.
    \item An just-in-time learning system based on RAG is realized, capable of acquiring knowledge instantly.
    \item Xiwu, the first LLM specilized for high energy physics outperforms the foundation model in accuracy for domain-specific knowledge question answering, and exceeds GPT-4 in BOSS (BESIII Offline Software System) code generation.
    \item The WebUI application has been deployed into HepAI platform, the related codes is open-sourced on GitHub.
\end{enumerate}

\section{Related Work}
\textbf{Language Models}$\quad$ The introduction of deep learning has greatly propelled the development of language models. Early applications of neural networks to language models, such as Feedforward Neural Networks\cite{bengio2000neural}, laid the groundwork. The emergence of RNN\cite{mikolov2012statistical, graves2013speech, kalchbrenner2013recurrent} and LSTM\cite{graves2012long} models advanced language model by effectively capturing temporal dependencies. However, their inability to process sequences in parallel limited their efficiency. In 2017, the introduction of the Transformer model\cite{vaswani2017attention}, with its self-attention mechanism, overcame these limitations, significantly boosting the handling of long-distance dependencies and computational efficiency, propelling the development of language models into a new era. Building on this architecture, BERT\cite{devlin2018bert} and OpenAI's GPT series\cite{radford2018improving, radford2019language} were developed for substantial advancements in NLP tasks through pre-training and fine-tuning. Subsequent works \cite{yang2019xlnet, liu2019roberta, raffel2020exploring, clark2020electra} further refined and enhanced these methodologies, leading to continuous performance improvements. In 2020, The release of GPT-3\cite{brown2020language}, a significantly larger laguage model, underscored the effectiveness of few-shot and zero-shot learning, prompting a significant trend toward model scaling. By 2022, the launch of ChatGPT\cite{ChatGPT2022} marked milestones in NLP, showcasing the potential of Large Language Models (LLMs) as efficient AI communicators. Moreover, the open-source model Vicuna\cite{zheng2023judging}, which is based on fine-tuning LLAMA\cite{touvron2023llama}, made a substantial contribution to the open-source community by achieving up to 90\% of GPT-4's\cite{achiam2023gpt} performance, sparking widespread research interest.\cite{ghosal2023flacuna, mukherjee2023feasibility, zhu2023minigpt, tang2023toolalpaca}.

\textbf{LLMs in Specific-Domain}$\quad$  Recent research has shown that BERT models, when fine-tuned for specific domains, can achieve significant performance improvements. This trend is exemplified by models such as BioBERT\cite{lee2020biobert}(targeted at the biomedical domain), SciBERT\cite{beltagy2019scibert} (for scientific literature), and ClinicalBERT\cite{alsentzer2019publicly}(for clinical data). However, the high costs associated with pre-training large-scale, domain-specific models like Galactica\cite{taylor2022galactica} and BloombergGPT\cite{wu2023bloomberggpt} underscore the value of domain-specific fine-tuning as a more practical and cost-effective strategy. For instance, Flan-PaLM\cite{singhal2022large} constructs a specialized large model for the clinical medicine domain by combining few-shot learning, chain-of-thought (CoT) prompting strategies, self-consistency prompting, and instruction fine-tuning. WizardMath\cite{luo2023wizardmath} fine-tunes LLAMA2\cite{touvron2023llama2} using the Reinforcement Learning from Evol-Instruct Feedback method to create a large model for mathematical reasoning. Meanwhile, SciGLM\cite{zhang2024sciglm} develops a specialized large model for the scientific domain by fine-tuning with high-quality datasets built using a self-reflexive annotation framework. These studies showcase the potential and value of fine-tuning large language models for specific domains. 

\textbf{LLMs in HEP} In the field of high-energy physics, ChATLAS has trained a domain-customized model using documents such as Twiki, ATLAS docs, and the E-group Archive, providing AI services to the collaboration group. AccGPT, developed based on LLAMA, is used for accelerator auxiliary control at CERN. Based on ChatGPT, the ATLAS Open Data Higgs Analysis Guide can assist individuals interested in physics analysis who are not specialized in high-energy physics. LLMs are being experimented with in various types of task of HEP.

\section{Methodology}
\label{sec: methodology}

\subsection{Overall Architecture}

\begin{figure}[!htbp]  
    \centering
    \includegraphics[width=0.7\textwidth]{"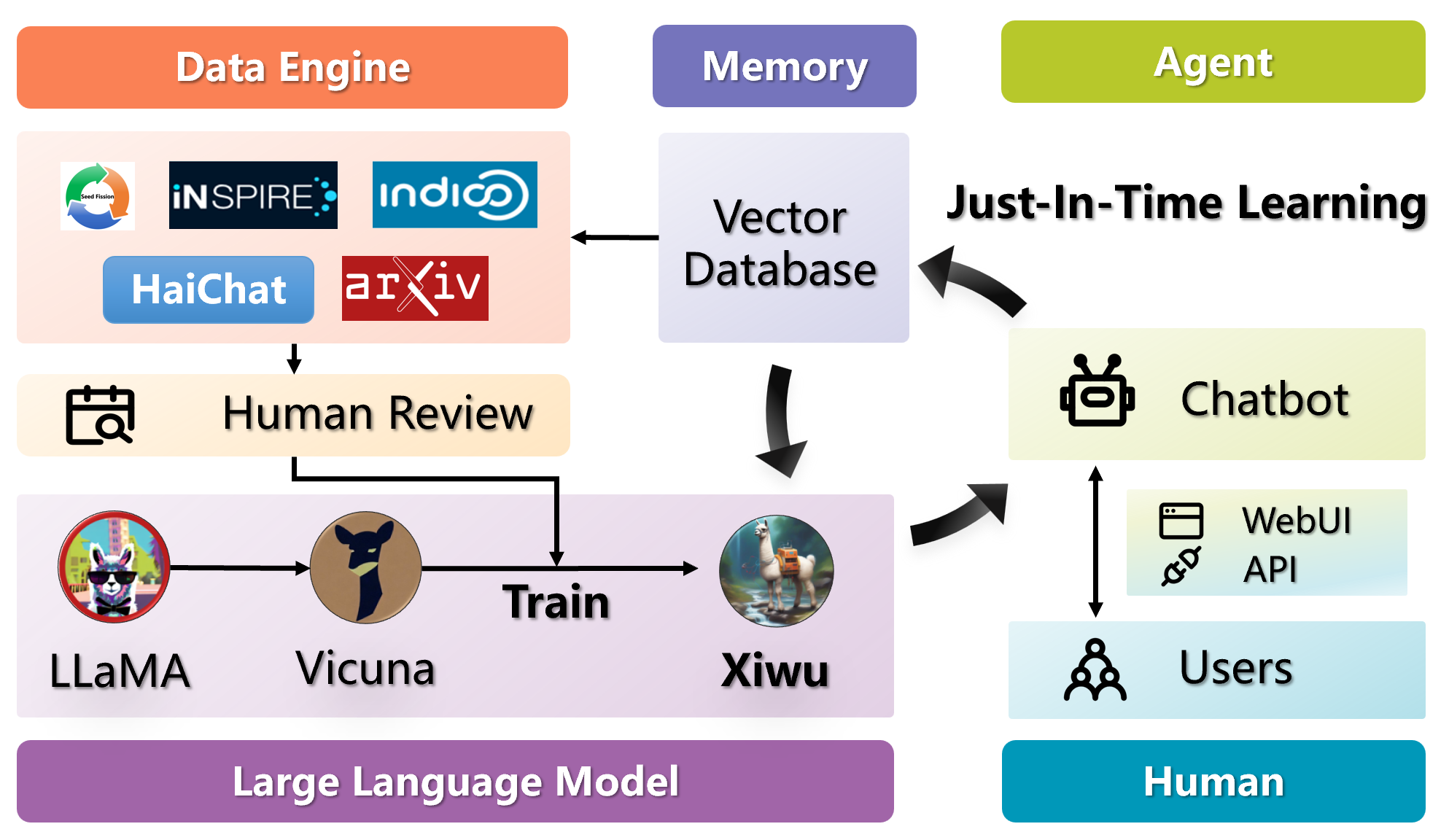"}
    \caption{The architecture of Xiwu large language model system.}
    \label{fig:arch}
\end{figure}

The architecture of Xiwu LLM is shown in FIG. \ref{fig:arch}. It consists of four parts: data engine, large language model, memory based on external knowledge lib and intelligent agent with its interface.
\paragraph{Data Engine} The data engine is primarily used to collect text data from eight fields related to HEP. The specific methods of data collection include seed fission technology, real concerns from ChatBot, knowledge from highly cited full-text literature, and a large number of literature abstracts. The data is cleaned and organized using LLM, and reviewed by professional staffs. 
\paragraph{Xiwu LLM} Based on the Level 0 Model LLaMA open-sourced by Meta, the Lmsys team has implemented a Level 1 Vicuna through secondary training. Currently, Xiwu is a Level 2 model trained on high-energy physics data based on Vicuna. With the upgrades of LLaMA and Vicucna, Xiwu's current base model is Vicuna-1.5 (corresponding to LLaMA2), with parameter sizes of 7B and 13B. Plans are in place to adapt to even more advanced models.
\paragraph{Memory Module} RAG has been widely used to mitigate hallucinations in LLMs, essentially enhancing the model's capabilities through external knowledge library. Similarly, we vectorize HEP data using an Embedding model and store it in a vector database as the long-term memory for Xiwu. Additionally, the process of adding or modifying information in the vector database represents a low-cost, rapid, and just-in-time learning process.
\paragraph{Intelligent Agent} The intelligent agent is structured around the Xiwu model, supplemented by an onboard knowledge base. Utilizing our proposed HepAI Distributed Deployment Framework (HepAI-DDP), we've successfully implemented both an Application Programming Interface (API) and a ChatBot Web User Interface (WebUI). This advancement significantly simplifies the interaction process between the intelligent agent and its human counterparts, be they scientists or developers.
\paragraph{Learning Loops} There are two learning loops in the system. The first loop is a just-in-time learning system, wherein users can directly embed accurate information into the knowledge base via the WebUI or API. Xiwu then retrieves and synthesizes this information to provide answers to queries. The second loop involves stable, unchanging information being integrated into the HEP text datasets, which are subsequently used to fine-tune the weights of Xiwu model. The former loop facilitates the rapid assimilation of evolving information, while the latter progressively enables the model to "understand" knowledge.

\subsection{Datasets}
\label{sec:dataset}

\begin{figure}[!htbp]  
    \centering
    \includegraphics[width=0.8\textwidth]{"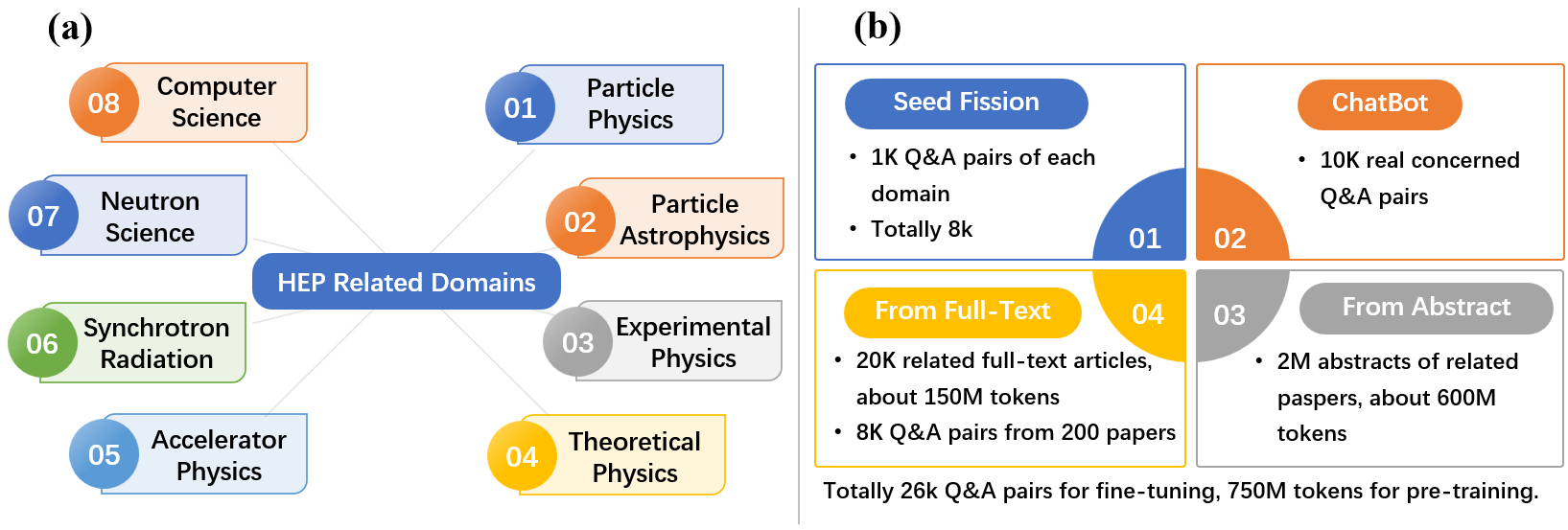"}
    \caption{The data resources and acquisition methods. (a) Eight domains related to High Energy Physics that are of our concern; (b) Four methods employed to gather the dataset}
    \label{fig:dataset}
\end{figure}

The FIG. \ref{fig:dataset} shows HEP text data from eight sub-domains and the data collection methods. As shown in FIG. \ref{fig:dataset} (a), in order to ensure the diversity and breadth of the model, we extensively gathered Q\&A pairs and literatures from eight HEP related domains, including particle physics, particle astrophysics, experimental physics, theoretical physics, accelerator physics, synchrotron radiation, neutron science and computer science. Data of each field is collected using the methods shown in FIG. \ref{fig:dataset} (b) on the right. 
There are four methods we used to collect data, including seed fission technology, chat robot backend, full-text of highly-cited papers and abundant abstracts of the papers. 
Through those methods, totally 26k Q\&A pairs for fine-tuning and 750M tokens for pre-training are collected and clearned up.\\
The details will be described below.

\subsubsection{Data from Seed Fission Technology}

\paragraph{Seed Fission} The seed fission technology was proposed by us to quickly obtain related Q\&A pairs in a specific domain. This technology allows a single seed such as \textit{Particle Physics}, can fission into a multitude of diverse and in-depth Q\&A pairs. 
By employing this technique, we successfully generated more than 1K Q\&A pairs for each HEP-related domain as shown in FIG. \ref{fig:dataset} (a). After a thorough human review, we obtained about 8K high-quality data entries.

\begin{figure}
    \centering
    \includegraphics[width=0.8\linewidth]{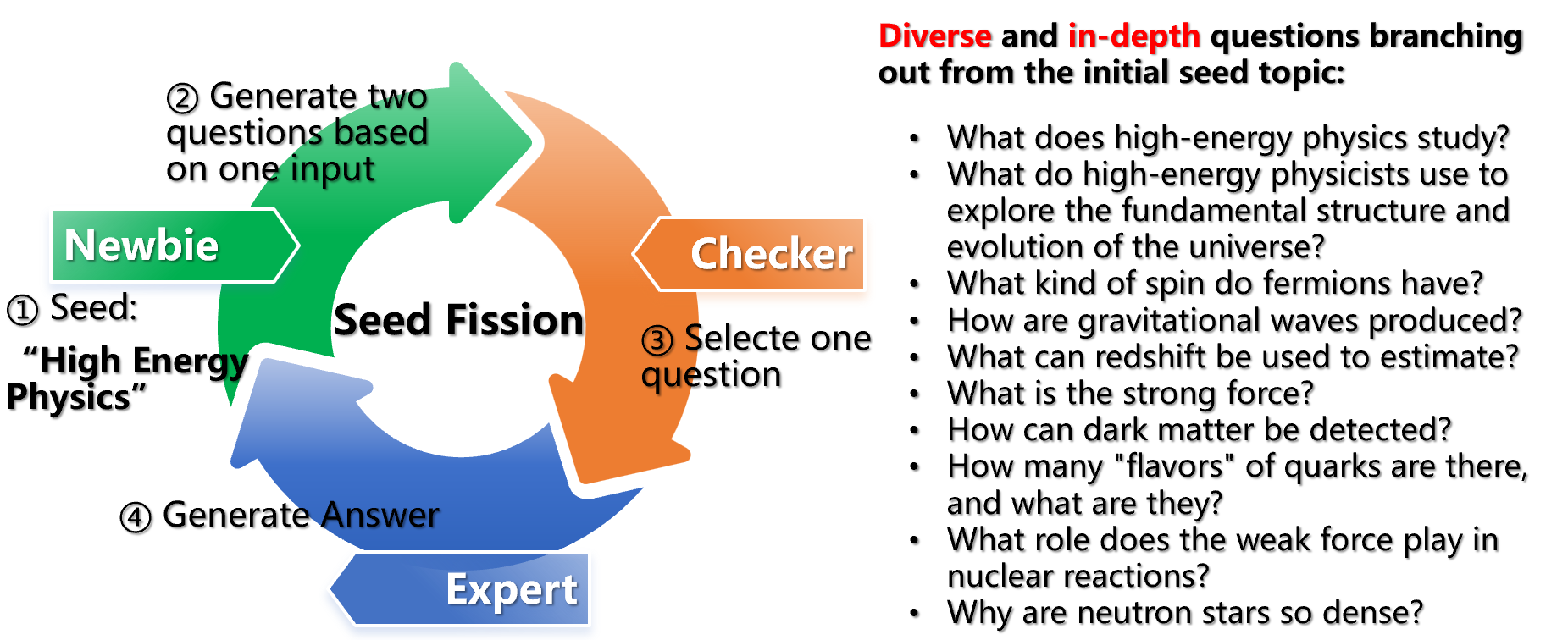}
    \caption{The seed fission technology for getting diverse and in-depth data}
    \label{fig:sf}
\end{figure}

\paragraph{Fission Process} As shown in FIG. \ref{fig:sf}, our approach involved designing three chat robots through prompt engineering: Newbie, Expert and Checker. The Newbie asking questions with curiosity based on information. The Expert providing answers based on information and reliable sources as much as possible. The Checker is a topic supervisor that responsible for selecting topics and determining when to transition away from the current topic. The fission process is as follows:
\begin{enumerate}
    \item The initial topic is considered a seed, such as "Particle Physics," is given to the Newbie. 
    \item The Newbie generates 2 to 10 random questions based on the input. A single input leading to multiple outputs is thus referred to as fission. 
    \item These questions are then reduplicated and filtered by the Checker, selecting the most interesting ones, or deciding to exit the current topic to choose another candidate topic. The selected question is directed to the next robot, while the remaining questions go into a candidate question list.
    \item The Expert answers questions based on the knowledge base and search engine it is equipped with, and provides the source of the answers.
    \item The answer serves as input information for the Newbie, who then poses more questions.
\end{enumerate} 

\paragraph{Data Quality} By using "Particle Physics" as the "seed," 50 question-answer pairs and 2,822 candidate questions can be generated. Some samples are shown in FIG. \ref{fig:sf}, from just this seed alone, concepts related to high-energy physics such as fermions, spin, gravitational waves, and redshift can be derived, thus the data generated is diverse and in-depth.
\paragraph{Advantage} The significant advantage of this technique is that it allows us to generate a large volume of relevant and diverse question-answer datasets with depth using just one topic as a guide. 
\paragraph{Limitations} Initially, a notable limitation is the tendency for both Newbie and Expert robots to become ensnared within similar topics. Subsequently, we introduced the Checker to oversee this process, effectively addressing the issue. Another limitation is that the quality of the generated answers depends on the robot itself, and there is a possibility of generating incorrect or even harmful information, equipping the Expert with a trustworthy knowledge base, knowledge retrieval capabilities, and incorporating human verification can to some extent solve this problem.

\subsubsection{Data from Chat Robot}
\paragraph{HaiChat Sever} We developed HaiChat~\cite{HaiChat2023} to provide generative model services for HEP researchers. The backend includes API for close models, local deployment of open-source models, and Xiwu model. High Energy Physics (HEP) researchers utilize the HaiChat service in their daily work, generating real questions of interest and corresponding answers. This serves as an important data source for further improving the performance of the Xiwu model.
\paragraph{Data and Quality} As of now, HaiChat has generated about 600,000 Q\&A pairs, of which 8\%, or 48,000, are related to high-energy physics. To ensure the quality and accuracy of the data, we employed a multi-step filtering and cleaning process. First, we utilized the ChatGPT model to automatically filter out irrelevant or inaccurate dialogues from the collected question-answer pairs. Then, we performed manual review and editing to further enhance the quality and usability of the data. 

\subsubsection{Data from Highly Cited Papers}
\paragraph{Data Sources} For the eight sub-domains mentioned above, We analyzed the citation data of papers over the past five years and used tools to download a carefully selected set of 20,000 papers.
\paragraph{HaiNougat PDF Parser} Based on the Nougat~\cite{blecher2023nougat} model, we trained a PDF parsing algorithm, HaiNougat~\cite{HaiNougat2024}, using a domain-specific dataset, which achieves more accurate parsing of formulas and tables of HEP.
\paragraph{Q\&A Pairs Generation and Human Verification} Based on the parsed data, we extracted the core contributions, viewpoints, and results of the papers through prompt engineering, obtaining relevant question-and-answer pairs. To ensure quality, these were reviewed by humans.
\paragraph{Data Collected} From those highly cited papers, about 8,000 Q\&A pairs and 150M tokens were collected.

\subsubsection{Abstracts Data from arXiv}
Kaggle offers comprehensive coverage of research papers across various domains, but it does not provide the full text. We downloaded approximately 2 million paper abstracts from the aforementioned eight related fields on Kaggle. The quality of these abstract data is very high, they can be directly used for the secondary pre-training of the Xiwu model.

\subsubsection{Final Dataset}
After data collection and cleaning, a AI-Ready dataset of HEP texts was ultimately formed that can be used for LLM training, which includes 750M tokens for secondary pre-training and 26k Q\&A pairs for fine-tuning.

\subsection{Neural Networks}
\label{sec:model}

\subsubsection{Model}
\paragraph{Basis Flexible Design} Xiwu is positioned as a Level 2 LLM. With the goal of minimizing costs while maintaining advanced capabilities, it has been designed to be adaptable to foundational models of Level 0 or Level 1, including LLaMA, Vicuna, Chat-GLM, Grok-1 and more in the future. This adaptability is implemented in the code repository through an 'apis' folder, where interfaces for any model have appropriate API adaptations. For more details, you can refer to the Github~\cite{Xiwu-github-2024} repository.

\paragraph{Current Foundation Model} The Vicuna~\cite{vicuna2023} model was chosen as the foundational model after a comparative analysis, as its overall performance surpassed that of others like LLaMA-13B, LLaMA2-13B, Koala-13B, Oasst-pythia-13B, Alpaca-13B, and ChatGLM-6B.
Vicuna is a Level 1 model based on LLAMA, and currently, we are using Vicuna-1.5, which corresponds to LLAMA-2. In this model, the transformer architecture is employed. Compared to Recurrent Neural Networks (RNNs), the transformer model has proven to be superior in quality for many sequence-to-sequence tasks while being more parallelizable.

\paragraph{Modeling of NLP Tasks} The language modeling task is to assign a probability for the likelihood of a given word (or a sequence of words) to follow a sequence of words. A sequence of tokens are passed to the embedding layer first, followed by a positional encoding layer to account for the order of the word. Positional Encoding module injects some information about the relative or absolute position of the tokens in the sequence. The BPE (Byte-Pair Encoding) is used, the size of vocabulary is 32000. The positional encoding, RoRE and Pre-LN (layer normalization)~\cite{ding2021cogview} is used..

The Pre RMS Norm is as follows:

\[RMS(x) = \sqrt{\frac{1}{N}\sum_{i=1}^N{x_i^2}}\]

The SwishGLU is used as activation function, there are tree trainable weight matrix in SwishGLU, the formular of SwishGLU is
\[SwishGLU=Swish_{\beta}(xW) \otimes xV\]

The $n_{heads}$ is 32, the $head_{dim}$ is 128, the $FFN_{dimension}$ is 11008, the $h_{dimension}$ is 4096.

\subsubsection{Training Techniques}

\begin{figure}[!htbp]  
    \centering
    \includegraphics[width=0.8\textwidth]{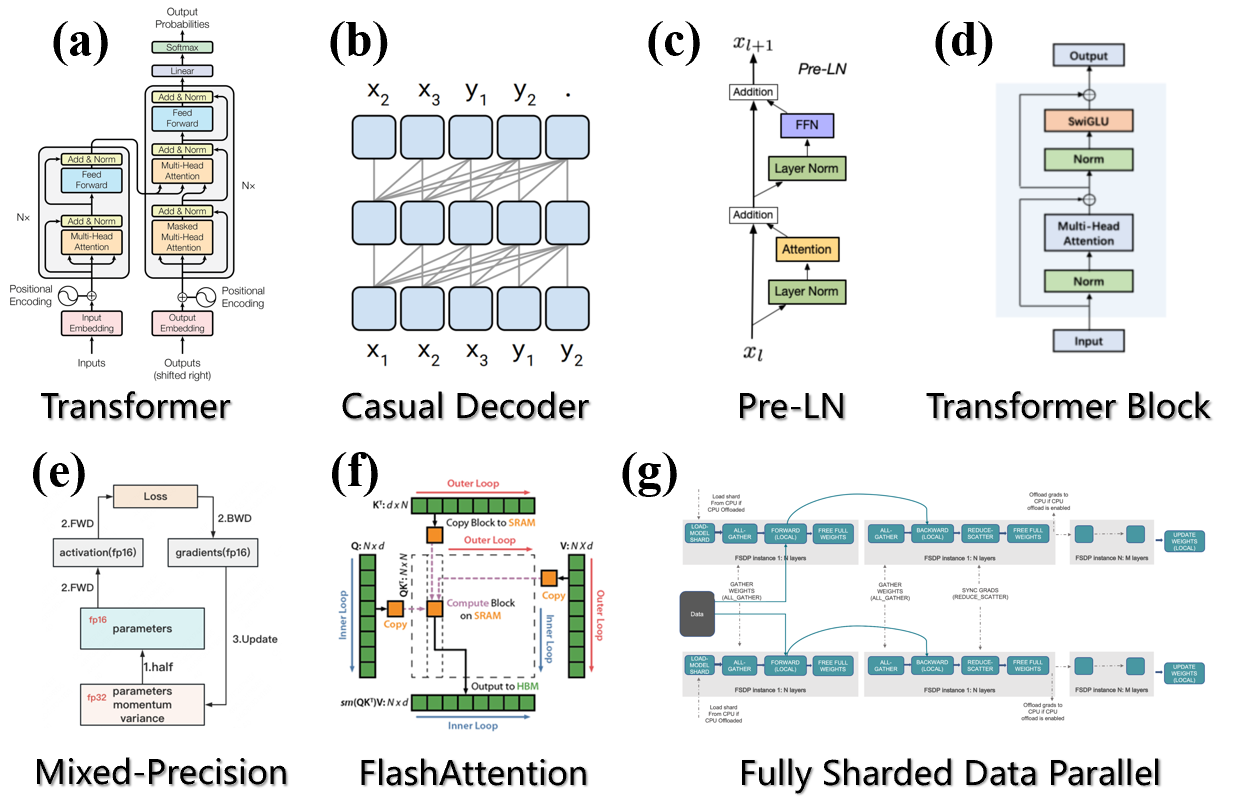}
    \caption{The illustration algorithm components and training technologies}
    \label{fig:compoents}
\end{figure}

During the training process, several advanced techniques are employed to enhance the performance and efficiency of the Xiwu model. These techniques, namely BF16 and TF32 mixed-precision training, FlashAttention, FSDP (Fully Sharded Data Parallelism), CPU-Offload, and Torchrun, played a crucial role in achieving improved training outcomes. In this section, we provide a detailed explanation of these techniques and their effects.

\paragraph{Mixed-precision Training} Mixed-precision training\cite{micikevicius2017mixed} involves performing computations using a combination of lower (BF16) and higher (TF32) precision formats. By utilizing BF16 for most of the computations and TF32 for certain critical operations, we were able to reduce memory usage, improve computational throughput, and accelerate the training process without sacrificing model accuracy.
\paragraph{FlashAttention} FlashAttention~\cite{dao2022flashattention} is a technique designed to enhance Flops (floating-point operations per second) utilization during self-attention computations, a critical component in transformer-based models like Xiwu. It optimizes the memory access patterns and computation flow, resulting in improved computational efficiency and reduced training time.
\paragraph{FSDP} Fully Sharded Data Parallelism (FSDP)\cite{zhao2023pytorch} is a data parallelism technique that enables efficient parallel training across multiple devices by dividing the model's parameters into shards and assigning each shard to a different device. By distributing the model's parameters, FSDP reduces memory consumption per device, enables larger models to fit within device memory constraints, and facilitates seamless parallelization for faster and more scalable training.
\paragraph{CPU-Offload} CPU-Offload\cite{pudipeddi2020training, rhu2016vdnn} is a strategy used to alleviate excessive GPU memory usage during training. By offloading certain computations to the CPU, we were able to free up GPU memory for more efficient memory utilization. This approach allowed us to train Xiwu on GPUs without encountering memory limitations and improved the overall training performance.
\paragraph{Torchrun for Single-Node Multi-GPU Parallelism} Torchrun is a utility that facilitates single-node multi-GPU parallelism in PyTorch\cite{paszke2019pytorch}, the framework used for developing Xiwu. By leveraging Torchrun, we achieved an effective distribution of computations across multiple GPUs within a single node, allowing for parallel training and accelerated model convergence.
\paragraph{Bitsandbytes} Bitsandbytes\cite{DettmersBitsAndBytes} is an innovative software designed to address the substantial GPU memory requirements encountered by LLMs during inference. Employing vector quantization and mixed-precision decomposition, Bitsandbytes substantially reduces memory consumption without compromising on performance. It enables the deployment of large-scale models, such as those with 175B parameters like OPT-175B/BLOOM\cite{le2022bloom}, on consumer-grade GPU servers, presenting a highly efficient solution for managing LLMs.
\paragraph{Deepspeed} The Deepspeed framework\cite{DeepSpeed} is designed to accelerate deep learning training by utilizing mixed precision, gradient checkpointing, and offloading memory to the host CPU. It allows for larger models and batch sizes, reducing memory requirements and improving training speed.
\paragraph{ZeRO Series} Zero Redundancy Optimizer series techniques aim to reduce memory consumption during training by partitioning model weights and optimizer states across multiple devices or nodes. These techniques include ZeRO-2\cite{rajbhandari2020zero}, ZeRO-3\cite{rajbhandari2020zero}, and ZeRO-offload\cite{ren2021zero}. By leveraging memory optimization strategies such as activation checkpointing and optimizer state partitioning, ZeRO techniques enable training larger models that would otherwise exceed available memory limits. Adopting ZeRO techniques can potentially allow us to scale up Xiwu and achieve even better performance.
\paragraph{LoRA} Low-Rank Adaptive technique\cite{hu2021lora} is a low-rank approximation technique that reduces the computational cost of large-scale language models. It achieves this by approximating the weight matrices with low-rank factors while preserving the model's representational capacity. By leveraging the inherent redundancy in model parameters, LoRA significantly reduces both the memory and compute requirements, making training and inference more efficient.

These techniques collectively contributed to the enhanced performance and efficiency of Xiwu during the training process. By leveraging mixed-precision training, FlashAttention, FSDP, CPU-Offload, and Torchrun, we achieved improved memory utilization, accelerated training speed, and scalable parallelization. These advancements enable us to train Xiwu more effectively, ultimately leading to improved performance in high-energy physics question-answering tasks.

\subsubsection{Computing Power and Training Settings}

\paragraph{Computing Power} We conducted full-scale training of the Xiwu-13B model using eight A100 GPUs, each with a memory capacity of 40 GB. Additionally, the Xiwu-7B model can be trained on two A100 GPUs with a memory capacity of 80 GB each.

\paragraph{Training Parameters} The following parameters were carefully selected and tuned to optimize the training process and improve the performance of Xiwu:
\begin{itemize}
    \item Learning rate: 2e-5
    \item Learning rate scheduler: cosine
    \item Weight decay: 0.0
    \item Warmup ratio: 0.03
    \item Per-device training batch size: 2
    \item Gradient accumulation steps: 16
\end{itemize}
\paragraph{Learning Rate} The learning rate determines the step size at which the model adjusts its weights during training. A value of 2e-5 was chosen to strike a balance between ensuring stable convergence and preventing overshooting or getting stuck in sub-optimal local minima.
\paragraph{Learning Rate Scheduler} The cosine scheduler gradually reduces the learning rate as training progresses. It helps the model to converge smoothly by reducing the learning rate towards the end of training. The cosine scheduler was selected due to its proven effectiveness in stabilizing training and improving the model's generalization ability.
\paragraph{Weight Decay} Weight decay is a regularization technique that adds a penalty to the loss function based on the magnitude of the model's weights. A weight decay value of 0.0 indicates that no additional regularization is applied during training, allowing the model to fully utilize the available information in the training data.
\paragraph{Warmup Ratio} The warmup ratio determines the proportion of training steps dedicated to gradually increasing the learning rate from zero to its initial value. A warm-up ratio of 0.03 ensures that the model's learning rate ramps up slowly at the beginning of training, allowing the model to stabilize and avoid sudden, large weight updates.
\paragraph{Per-device Training Batch Size} This parameter determines the number of samples processed in parallel on each device during training. A batch size of 2 per device strikes a balance between utilizing the available memory efficiently and maintaining reasonable training speed.
\paragraph{Gradient Accumulation Steps} Gradient accumulation is a technique that helps simulate larger batch sizes by accumulating gradients over multiple smaller batches before performing weight updates. With 16 gradient accumulation steps, the model benefits from an effective utilization of computational resources while still obtaining accurate gradient estimates.

It is important to note that the presented configuration and training setup can be adjusted according to the available resources and specific requirements of different applications.

\subsection{Just-In-Time Learning System}
\label{sec:memory}

\paragraph{Limitations of Secondary Training} Although secondary training of LLMs is effective, it is costly, inefficient, and training can lead to a degradation in model performance. For example, when dealing with unseen information, the model cannot effectively learn new information under the usual learning intensity (lower learning rates and fewer epochs). If the learning intensity is increased, the model can learn new information, but then issues such as performance degradation on other Q\&A tasks and repetitive answer fragments may occur.

\paragraph{Just-In-Time Learning} Just-In-Time Learning (JITL) is a learning system that utilizes a vector database for long-term memory enhancement to augment model capabilities, operating in a basic process equivalent to RAG. We call it JITL to emphasize its learning attributes. JITL is capable of rapid knowledge learning in seconds, specific knowledge updating, and unloading. These features are not present in the traditional model training approach.

\begin{figure}[!htbp]  
    \centering
    \includegraphics[width=0.9\textwidth]{"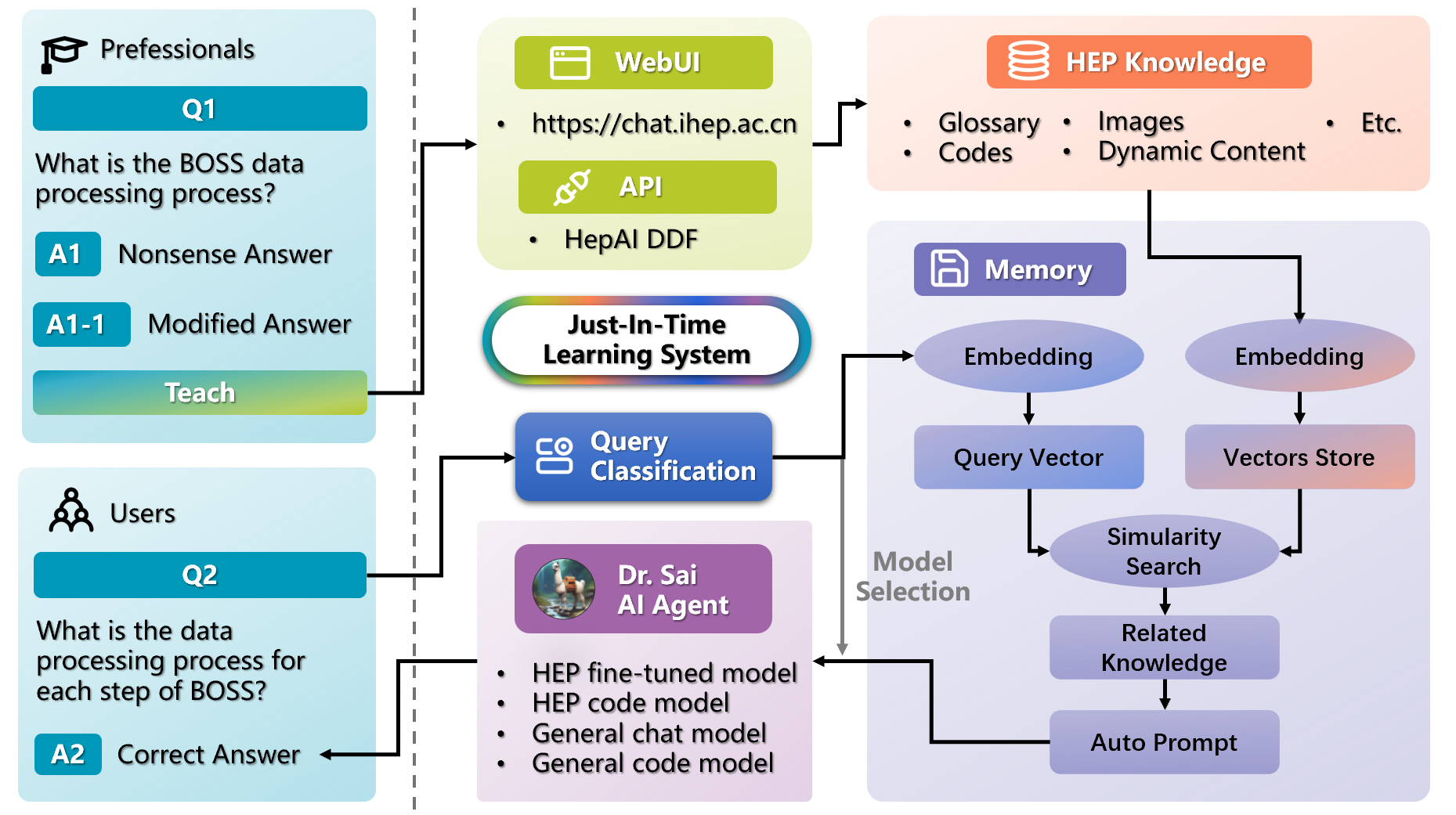"}
    \caption{The just-in-time learning system}
    \label{fig:jitl}
\end{figure}

\paragraph{Implementation of JITL} The architecture of the Just-in-time Learning system is illustrated in FIG. \ref{fig:jitl}, where the core concept is integrating an external knowledge base as the memory part of the model. New information is saved in the knowledge base, and after retrieval, comprehensive information is outputted. The complete information flow of the JITL system is as follows:
\begin{enumerate}
  \item When an expert user asks the model a specialized question related to HEP and receives a nonsensical answer, the expert can suggest modifications or directly edit the answer, then teach the model via WebUI or API.
  \item The system categorizes and organizes the new information, storing it in a text database.
  \item The text data is processed by a pre-trained embedding model to generate feature vectors, which are then stored in a vector database. These feature vectors capture the semantic information of the text, and semantically similar expressions are close to each other in the feature space.
  \item When ordinary users ask the model questions, these go through a question classifier and select an appropriate model, while text data is converted into feature vectors by the embedding model. As different texts are separable in high-dimensional vector space, a simple similarity search can retrieve stored related texts.
  \item The related texts and query are processed through automatic prompt engineering and then fed into a suitable LLM such as Xiwu. Due to the intent understanding and contextual learning abilities of Xiwu, the final output is an accurate answer that also includes information sources.
\end{enumerate}

\paragraph{Advantages} Compared to training models, Just-In-Time learning systems only store vectors after embedding into the database, making the learning process extremely fast, achieving learning in seconds. Knowledge can be dynamically updated, which is particularly effective for rapidly changing information. Knowledge can also be specifically unloaded, a feature not achievable in traditional model training processes. Compared to conventional retrieval systems, Just-In-Time learning systems accomplish semantic-level rather than keyword-level retrieval, resulting in substantial performance improvements.

\paragraph{Limitations} Just-In-Time learning systems have limitations. First, the additional vector database search process can be time-consuming, leading to higher latency compared to pure LLMs. One mitigation strategy is to use GPU acceleration to speed up the search process, another approach is to train the model with information that is always correct. The second limitation is that the model doesn't truly "understand" the knowledge in the external database, meaning there's no improvement in reasoning ability. This mode is akin to open-book exams where students don't have to understand the material but can look up information when needed. 

\section{Evaluation}
\label{sec:evaluation}

\subsection{Evaluation between Models}

\paragraph{Evaluation Data} To assess the performance of Xiwu in comparison to other models, we conducted a human evaluation using a set of 5,100 prompts. These prompts consisted of questions related to particle physics, astrophysics, synchrotron physics, and neutron science. Examples of these questions include: "What quarks make up protons and neutrons?", "What is general relativity?", "Why can't we directly observe dark matter?", and "Why is particle therapy more expensive than X-ray treatment?". The evaluation dimensions included accuracy, clarity, and fluency. Accuracy assessed whether the answers correctly addressed the questions and contained the necessary information. Clarity evaluated the clarity and comprehensibility of the answers, including the use of accessible language and terminology. Fluency assessed the overall fluency and coherence of the answers. Human scientists participated in the evaluation process and compared the answers generated by Xiwu with those from other models, namely Vicuna-13B and ChatGPT. The evaluation results were categorized into three outcomes: Xiwu wins, Xiwu fails, and Draw.

\begin{figure}[!htbp]
    \centering
    \includegraphics[width=0.5\textwidth]{"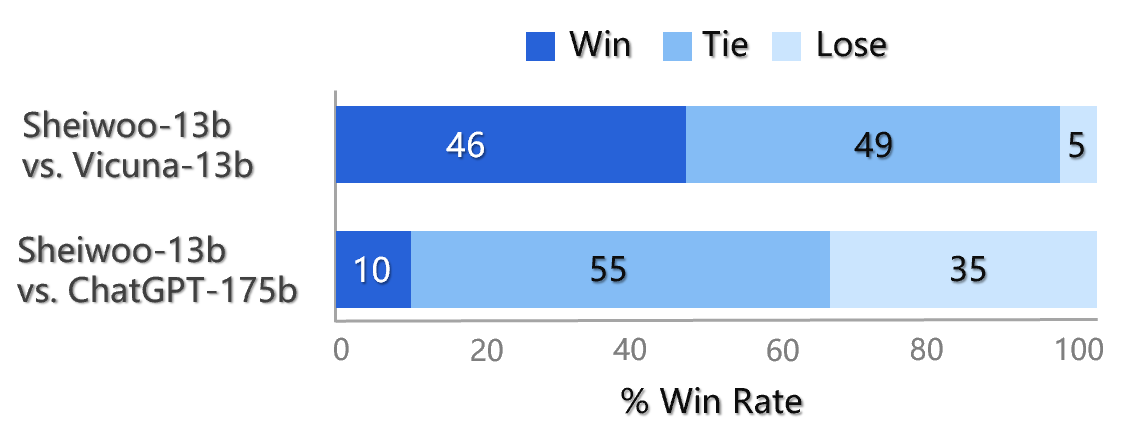"}
    \caption{Human preference evaluation, Compare Xiwu-13B with Vicuna-13B and ChatGPT-175B}
    \label{fig:winrate}
\end{figure}

\paragraph{Comparison Results} As shown in the FIG. \ref{fig:winrate}, Xiwu-13B achieved a win or draw rate of 95\% when compared to the baseline model Vicuna-13B. The performance of Xiwu-13B reached 65\% of ChatGPT-175B, while Vicuna achieved less than 10\%. These findings indicate that fine-tuning a large-scale model with domain-specific data, even on a model with 13 billion parameters, can significantly improve the accuracy of domain-specific question answering. Fine-tuning smaller models based on general-domain knowledge has the potential to achieve or even surpass the performance of larger models in specialized domains.

\paragraph{Fine-tuning Cannot Learn New Knowledge} Furthermore, we also tested the ability of Xiwu to incorporate internal knowledge that was not pre-trained. Examples of such knowledge included "What is HepAI?" (a high-energy physics artificial intelligence platform we are developing) and "What is HaiGF?" (an AI application interface framework we are developing). The results showed that a limited number of epochs were insufficient to guide Xiwu in answering these types of questions. Increasing the number of epochs not only failed to yield correct answers but also led to a decrease in the model's performance on other questions.

\subsection{Absolute Evaluation}

\paragraph{Results} In the absolute evaluation, human scientists assessed the content of each Xiwu answer, categorizing them as Excellent (Xiwu provided an outstanding response), Pass (Xiwu's answer was acceptable), or Fail (Xiwu's answer was incorrect). During the absolute evaluation, it was observed that occasionally the 13-billion-parameter Xiwu model produced answers that did not understand the human intent, resulting in responses that were unrelated to the questions. Additionally, there were instances where the model repeated the same text. This behavior may be attributed to the lack of reinforcement learning from human feedback. Further investigation and incorporating reinforcement learning techniques are necessary to address these issues.

\section{Conclusion}
\label{sec:conclusion}

\paragraph{First LLM for HEP} In this work, we developed Xiwu, the first large language model customized for high energy physics. Xiwu features a flexible foundational model and two distinct learning systems. The flexibility of the foundational model allows Xiwu to evolve alongside the development of open-source models. The learning system based on model training is traditional, enabling the model to effectively learn and "understand" HEP knowledge. The vector database-based just-in-time learning system enables rapid learning of new knowledge, dynamic knowledge updates, and unloading, all at a low cost. 
\paragraph{Results} Currently, Xiwu is an Level 2 LLM based on Vicuna. Xiwu-13B significantly outperforms the Vicuna-13B on the HEP domain Q\&A test set, achieving about 65\% of the performance of ChatGPT-175B. Furthermore, our just-in-time learning system enables multiple individuals to collectively teach Xiwu, demonstrating the potential for collaborative AI training. The seed fission technology we have developed has potential for wide application, capable of generating Q\&A pair data in diverse fields.
\paragraph{Meaning} This paper not only presents our findings but also provides the corresponding implementation code, which is readily available on GitHub~\cite{Xiwu-github-2024}. We hope that our work will inspire further research and development in the application of large language models in specialized scientific fields.

\section*{Acknowledgments}
This work is Supported by the Informatization Plan of Chinese Academy of Science, Grant
No. CAS-WX2022SF-0104 and "From 0 to 1" Original Innovation Project of IHEP, Grant No. E3545PU2.
We would like to express our gratitude to Beijiang Liu, Yaquan Fang, Gang Li, Wuming Luo, Ye Yuan, Shengsen Sun, Yi Jiao and others who are not listed here for engaging in beneficial discussions or providing computing resources.

\bibliographystyle{unsrt}  
\bibliography{references}  

\end{document}